\documentclass[prl,aps,twocolumn,superscriptaddress,showpacs,nofootinbib]{revtex4-1}

\usepackage{graphicx}
\usepackage{textcomp}
\usepackage{color}
\usepackage[colorlinks]{hyperref}
\hypersetup{
	colorlinks = true,
	citecolor = {blue},
	linkcolor = {blue},
	urlcolor = {blue}
}

\begin{document}

\title{Non-line-of-sight imaging with picosecond temporal resolution}

\author{Bin Wang}
\thanks{Bin Wang and Ming-Yang Zheng contributed equally to this work.}
\affiliation{Hefei National Laboratory for Physical Sciences at the Microscale and Department of Modern Physics, University of Science and Technology of China, Hefei 230026, China}
\affiliation{Jinan Institute of Quantum Technology, Jinan 250101, China}

\author{Ming-Yang Zheng}
\thanks{Bin Wang and Ming-Yang Zheng contributed equally to this work.}
\affiliation{Jinan Institute of Quantum Technology, Jinan 250101, China}

\author{Jin-Jian Han}
\author{Xin Huang}
\affiliation{Hefei National Laboratory for Physical Sciences at the Microscale and Department of Modern Physics, University of Science and Technology of China, Hefei 230026, China}
\affiliation{Shanghai Branch, CAS Center for Excellence in Quantum Information and Quantum Physics, University of Science and Technology of China, Shanghai 201315, China}
\affiliation{Shanghai Research Center for Quantum Sciences, Shanghai 201315, China}

\author{Xiu-Ping Xie}
\affiliation{Jinan Institute of Quantum Technology, Jinan 250101, China}

\author{Feihu Xu}
\email[]{feihuxu@ustc.edu.cn}
\author{Qiang Zhang}
\email[]{qiangzh@ustc.edu.cn}
\affiliation{Hefei National Laboratory for Physical Sciences at the Microscale and Department of Modern Physics, University of Science and Technology of China, Hefei 230026, China}
\affiliation{Shanghai Branch, CAS Center for Excellence in Quantum Information and Quantum Physics, University of Science and Technology of China, Shanghai 201315, China}
\affiliation{Shanghai Research Center for Quantum Sciences, Shanghai 201315, China}
\affiliation{Key Laboratory of Space Active Opto-electronics Technology, Chinese Academy of Sciences, China}

\author{Jian-Wei Pan}
\email[]{pan@ustc.edu.cn}
\affiliation{Hefei National Laboratory for Physical Sciences at the Microscale and Department of Modern Physics, University of Science and Technology of China, Hefei 230026, China}
\affiliation{Shanghai Branch, CAS Center for Excellence in Quantum Information and Quantum Physics, University of Science and Technology of China, Shanghai 201315, China}
\affiliation{Shanghai Research Center for Quantum Sciences, Shanghai 201315, China}

\date{\today}

\begin{abstract}
	Non-line-of-sight (NLOS) imaging enables monitoring around corners and is promising for diverse applications. The resolution of transient NLOS imaging is limited to a centimeter scale, mainly by the temporal resolution of the detectors. Here, we construct an up-conversion single-photon detector with a high temporal resolution of $\sim 1.4$ ps and a low noise count rate of 5 counts per second (cps). Notably, the detector operates at room temperature, near-infrared wavelength. Using this detector, we demonstrate high-resolution and low-noise NLOS imaging. Our system can provide a $180$ \textmu m axial resolution and a $2$ mm lateral resolution, which is more than one order of magnitude better than that in previous experiments. These results open avenues for high-resolution NLOS imaging techniques in relevant applications.
\end{abstract}

\maketitle

\emph{Introduction.---}
Non-line-of-sight imaging can visualize hidden objects outside the direct field of view (FOV) of the camera. It has diverse applications in robotics, manufacturing, medical imaging, and exploration and rescue in extreme conditions, etc. In recent years, NLOS imaging has attracted growing interest \cite{Altmann2018Quantum,Tomohiro2019Recent,Faccio2020Review}. In contrast to conventional line-of-sight imaging \cite{Sun20133D,Kirmani2014First}, NLOS imaging analyses a small number of photons reflected from multiple surfaces, not coming directly to a photodetector. Consequently, this process involves several challenges in both theory and experiment \cite{Faccio2020Review}, such as algorithms for image reconstruction, high-efficiency single-photon detection, and high system temporal resolution for time-of-flight (TOF) measurements. Most of the optical NLOS imaging approaches rely on transient techniques, which use the information encoded in the TOF of photons scattered multiple times and employ high-resolution time-resolved single-photon detectors (SPDs) to detect the photons \cite{Kirmani2009NLOS,Velten2012StreakCamera,Raskar2012OE,Velten2014BP,Heide2014TOFcamera,Faccio2016Tracking, Raskar2016TOFcamera, buttafava2015gatedSPAD, O'Toole2018LCT, Faccio2019PRA, O'Toole2019f-k, Velten2019PhasorCamera, Xin2019FermatPath}. This technique has been moved forward from laboratory verifications to long-range demonstrations \cite{Faccio2017long-range,wu2021}. Furthermore, one may employ an ordinary camera with no TOF information to compute the NLOS images, but the reconstructed images usually do not involve depth information \cite{Katz2014Speckle,Klein2016TrackingCamera,Bouman2017,Xu2018Anti-pinhole,O'Toole2018Speckle-Tracking,Vivek2019RGBcamera,Heide2020CNN}.

The fundamental resolution of transient NLOS imaging is determined by the temporal resolution of the experimental setup \cite{O'Toole2018LCT}. In previous experiments \cite{Kirmani2009NLOS,Velten2012StreakCamera,Raskar2012OE,Velten2014BP,Heide2014TOFcamera,Faccio2016Tracking, Raskar2016TOFcamera, buttafava2015gatedSPAD, Faccio2017long-range, O'Toole2018LCT, Faccio2019PRA, O'Toole2019f-k, Velten2019PhasorCamera,Xin2019FermatPath}, the best achieved temporal resolution was in the range of tens of picoseconds, mainly limited by the underlying SPDs. As a result, the resolution of NLOS imaging was restricted to a centimeter scale. Note that several types of SPDs ranging from visible band to infrared band have been developed \cite{Hadfield2009NP,Zhang2015InGaAs, Marsili2013SNSPD, Robert2012SNSPD, Franco2004UCSPD, Fejer2011long-wavelength-pump}; however, most of these SPDs suffer from a low temporal resolution of dozens of picoseconds or more. This has become a bottleneck for high-resolution NLOS imaging.

Based on short laser pulse pumping technology \cite{Poole1998psUCSPD,Franco2008fsUCSPD,Zeng2012UCSPD,Delteil2019ps-g2,Rehain2020psImaging}, we construct an up-conversion single-photon detector (UCSPD) working at the near-infrared wavelength with a temporal resolution of $\sim 1.4$ ps. Compared with a superconducting nanowire single-photon detector having a picosecond temporal resolution \cite{korzh2020psSNSPD}, our detector works at room temperature without the need for cryogenic devices. A low noise counts rate of 5 cps is realized by long-wavelength-pumping \cite{Fejer2011long-wavelength-pump,Ma2017UCSPD1064} and time-gating \cite{supp} methods. With this UCSPD as the core detection device, we demonstrate a $180$ \textmu m axial resolution and a $2$ mm lateral resolution capability for NLOS imaging. The resolution is more than one order of magnitude better than that in previous NLOS experiments \cite{O'Toole2018LCT,O'Toole2019f-k,Velten2019PhasorCamera}. Furthermore, a precise confocal setup can be realized where the up-conversion process serves as an exceptional spectral-temporal filter to rejecting the unwanted noise \cite{Rehain2020psImaging} and offering an efficient way to remove the first bounce photons in NLOS imaging. These results open avenues for the development of high-resolution NLOS imaging techniques for relevant applications.

\begin{figure}[htbp!]
	\centering
	\includegraphics[width=\linewidth]{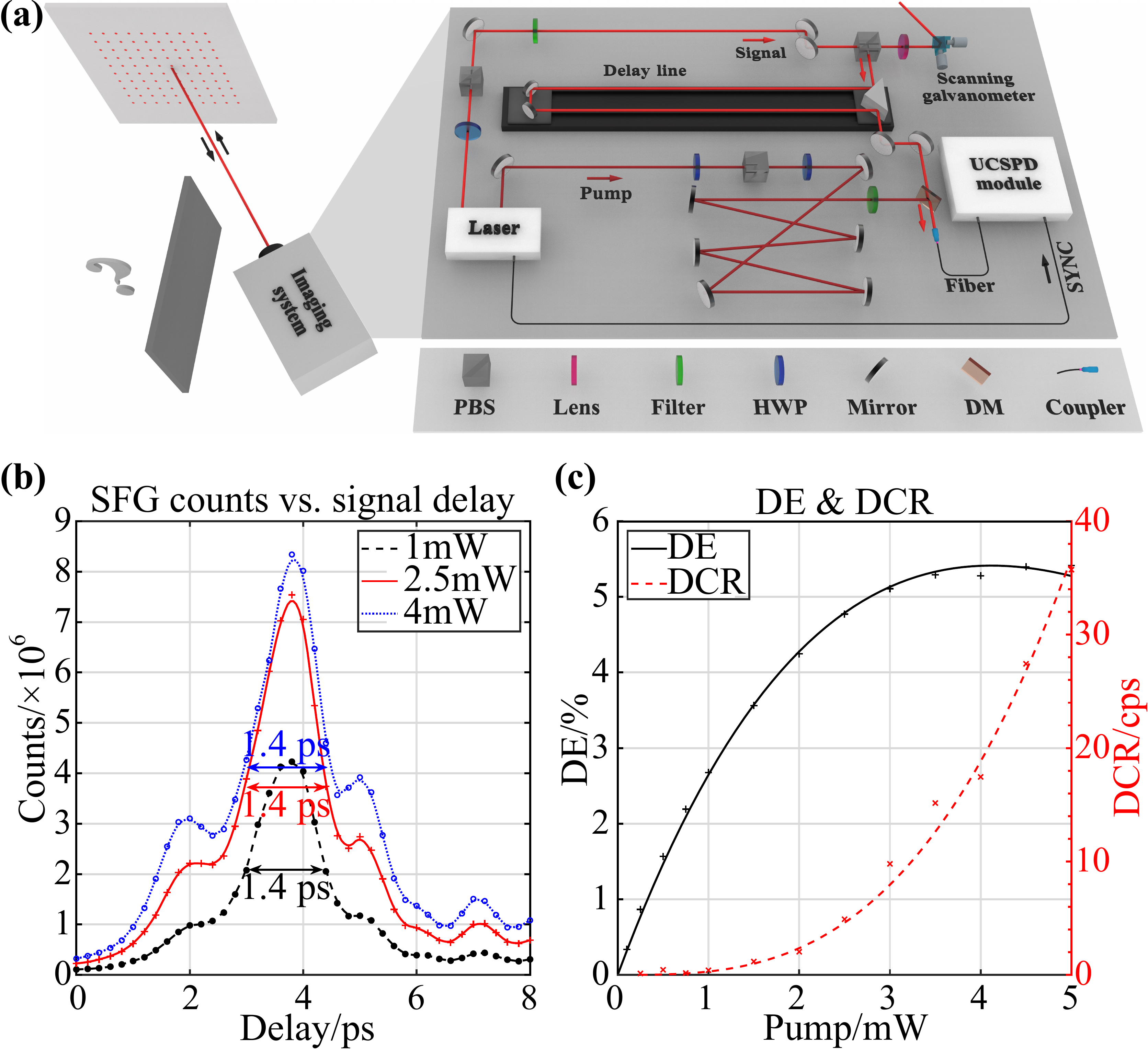}
	\caption{Experimental setup and testing of performance. (a) Experimental setup for NLOS imaging with a picosecond temporal resolution UCSPD. (b) The SFG photon counts while scanning the delay line with different pump power. (c) Detection efficiency and dark count rate of the SPD after time-gating on the TDC. The detection efficiency is calculated by dividing the maximum SFG count by the input signal photon count. }
	\label{fig:setup}
\end{figure}

\emph{Experimental setup.---}
The coaxial transient NLOS imaging system (Fig.~\ref{fig:setup}(a)) is based on the developed UCSPD with picosecond temporal resolution. A precisely synchronized two-channel laser emits two polarized picosecond laser pulses at $1053$ nm (signal) and $1550$ nm (pump) with a time jitter of dozens of femtoseconds between the two laser channels (see Section 1 in the Supplemental Material \cite{supp}). The full widths at half maximum (FWHM) of the two pulse lasers are $1.4$ ps and $1$ ps, respectively (see Fig. S1 in the Supplemental Material \cite{supp}). Both the signal and pump lasers work at a $60$ MHz repetition rate. A half-wave plate (HWP) together with a polarization beam splitter (PBS) provides the laser power tuning. The signal laser is focused on the visible wall with an $f = 500$ mm lens, and the illuminating point is raster-scanned with a 2-axis scanning galvanometer. The illumination spot and the FOV diameter are approximately $1.2$ mm and $0.4$ mm, respectively. The echo signal scattered from the visible wall and the hidden objects are transmitted by a fiber after passing a delay line. The pump laser is coupled with the signal laser path with a dichroic mirror (DM). A polarization controller (PC) is used for tuning the signal photon polarization. The signal photons and pump laser are then launched into a customized periodically poled lithium niobate (PPLN) waveguide (working at $ 20 $ $ ^{\circ} $C), and the sum-frequency generation (SFG) photons are detected by a silicon avalanche photodiode (APD) after noise filtering in the UCSPD module.

With a bit of rearrangement of the signal optical path, a weak signal laser can be transmitted directly into the UCSPD module to test the UCSPD temporal resolution (see Fig. S1(a) in the Supplemental Material \cite{supp}). By scanning the delay line with a $0.2$ ps increment, the obtained SFG photon counts show the correlation of the signal and pump pulse intensities depicted in Fig.~\ref{fig:setup}(b). The measured FWHM width is approximately $1.4$ ps while the theoretical minimum timing resolution is $1.4$ ps (see Section 1 in the Supplemental Material \cite{supp}). As pump power rises, the ratio of the primary peak to the secondary peak decreases due to the pump saturation (see Section 1 in the Supplemental Material \cite{supp}). By synchronizing the time-to-digital converter (TDC) with the pump laser and carrying out a time-correlated single-photon counting (TCSPC) measurement, a peak corresponding to the SFG photons will be formed  on the TDC histogram with a fixed time delay. Photon counts outside the SFG photons arriving time window are mostly the dark counts of the silicon APD. Therefore, a time-gating method can be adopted with this TDC (see Section 2 in the Supplemental Material \cite{supp}) to avoid dark counts of the silicon APD. The detection efficiency (DE) and dark count rate (DCR) of the SPD with picosecond temporal resolution is calculated after time-gating on the TDC (Fig.~\ref{fig:setup}(c)). Finally, we adopt a $2.5$ mW pump power accounting for the trade-off between conversion efficiency and pump saturation. An extremely low noise count rate of $ 5 $ cps under a $ 2.5 $ mW pump power after time-gating on the TDC is obtained compared to the previous work \cite{Zeng2012UCSPD,Delteil2019ps-g2,Rehain2020psImaging}.

Based on the picosecond temporal resolution UCSPD, the detection events will be time-stamped with a temporal resolution of $\sim 1.4$ ps. Furthermore, making possible the \emph{separate} detection (in the temporal domain) of the backscattered photons from the visible wall and the hidden targets on the delay line, a different illuminating power for the detection of the visible wall and the hidden targets can be applied by rotating the HWP in the signal path. This excludes the possible saturation on the silicon APD from the first bounce photons. Such flexibility also enables a more precise coaxiality than that in the standard confocal NLOS imaging system with a conventional SPD \cite{O'Toole2018LCT} and excludes the adverse impact of first bouncing photons. The deviation between the centers of the illuminating point and the FOV in our system is less than $200$ \textmu m. The $300$ \textmu W and $470$ mW illumination powers are applied to the wall for the detection of photons that are bounced back from the visible wall and the hidden targets, respectively.

Despite the reduced pump noise of $5$ cps with the time-gating on the TDC while the signal laser is off, $10^{2}$ - $10^{3}$ cps noise counts are detected on the whole range of the delay line with a $470$ mW signal laser illuminating the wall. They are caused by the continuous-wave amplified spontaneous emission (ASE) noise of the fiber amplified laser. Although the ASE noise is extremely weak compared to the signal pulses, the direct bounce of the ASE noise from the visible wall is remarkable compared with the echo photons from the hidden objects. Such ASE noise makes it difficult to obtain a high signal-to-noise ratio (SNR) using Lambertian hidden targets for reconstruction. Therefore, retro-reflective targets are used in all the imaging experiments.

\emph{Results.---}
The distance from the imaging system to the visible wall in our NLOS setup is $45$ cm, while the distance between the hidden objects and the wall is $18$ cm. $64 \times 64$ or $128 \times 128$ locations over a $6.4$ cm $\times$ $6.4$ cm area are sampled on the visible wall. Assuming that $\gamma$ is the temporal resolution of the imaging system, $2w$ is the width of the sampling area on the visible wall, and $z$ is the distance between the wall and the hidden objects, the theoretical axial resolution limit of $\Delta z = c\gamma/2$ and the lateral resolution limit of $\Delta x = c\gamma\sqrt{w^2 + z^2}/2w = \sqrt{1 + (z/w)^2}\Delta z$ \cite{O'Toole2018LCT}.  The optimal system temporal resolution of $1.4$ ps yields an ideal axial resolution limit of $210$ \textmu m and a lateral resolution limit of $1.2$ mm.

\begin{figure}[htbp]
	\centering
	\includegraphics[width=\linewidth]{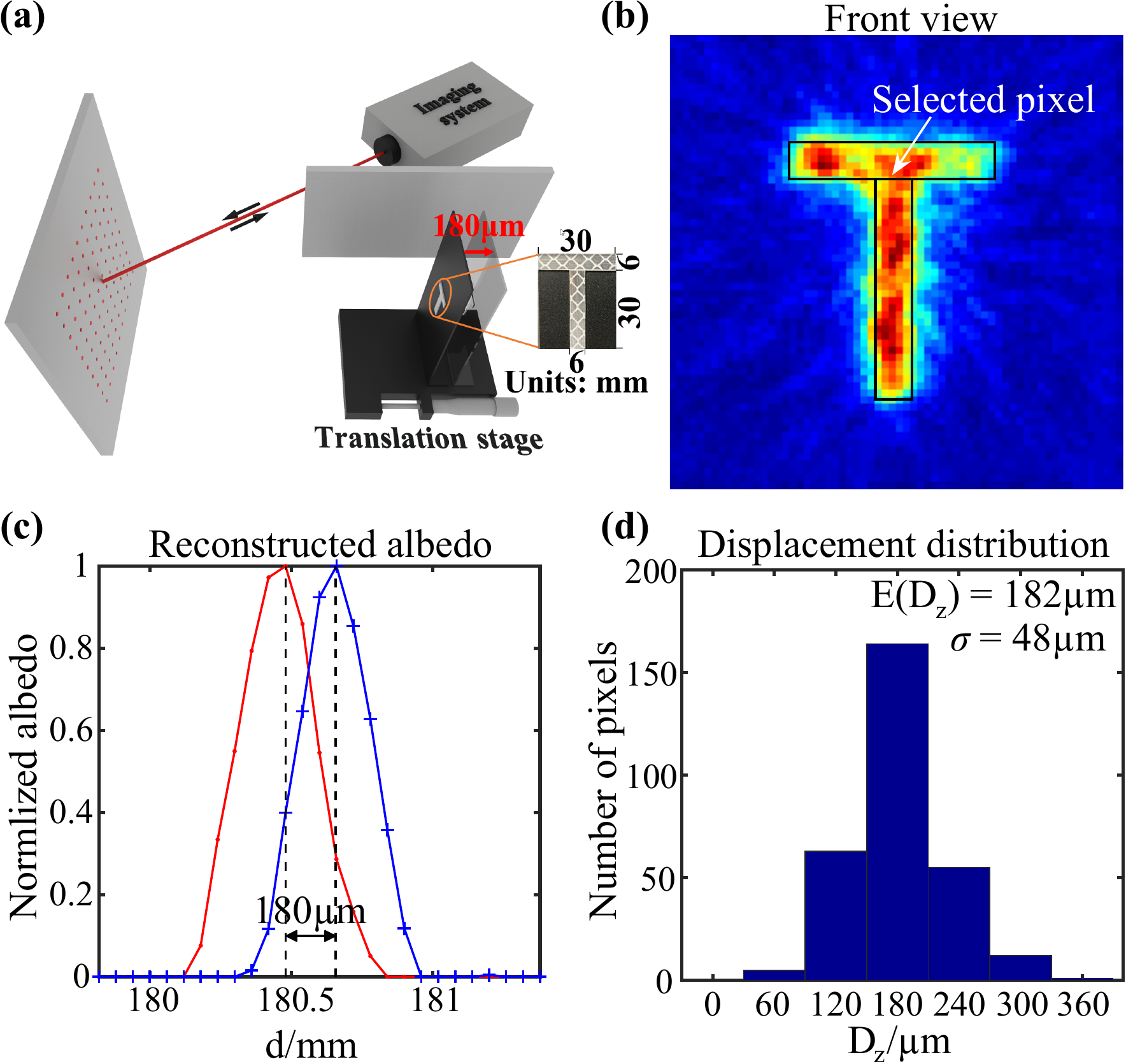}
	\caption{Determination of axial spatial resolution. (a) Experimental setup for axial resolution test. (b) One of the reconstructed images. The black box contains $ 300 $ pixels used in statistical analysis. (c) The reconstructed albedo along the axial direction of the selected pixel in Fig. 2(b). (d) Reconstructed distance distribution of the selected $ 300 $ pixels in Fig. 2(b).}
	\label{fig:axial_resolution}
\end{figure}

The procedure of determining the limit of axial resolution is illustrated in Fig.~\ref{fig:axial_resolution}. A hidden object in the form of $30$ mm $\times$ $36$ mm size retro-reflective letter 'T' is placed on a translation stage. Then a $64 \times 64$ grid sampling with a $1$ mm interval between adjacent sampling points on the visible wall is performed with the delay line moving with a step of $60$ \textmu m (corresponding to $0.4$ ps) and dwelling for $24$ ms on each time bin. After the first measurement, the stage is moved 180 \textmu m away from the wall to perform a second measurement. The light-cone transform (LCT) algorithm \cite{O'Toole2018LCT} was used for reconstruction. The axial position difference between two measurements of the pixel at the center of the reconstructed image (Fig.~\ref{fig:axial_resolution}(b)) is shown in Fig.~\ref{fig:axial_resolution}(c). The actual displacement of $180$ \textmu m is precisely reconstructed. There are $300$ pixels inside the black box in Fig.~\ref{fig:axial_resolution}(b), and data of those pixels from the two measurements are selected for the analysis of the displacement of the maximum albedo position. The obtained mean axial position displacement of $182$ \textmu m with a standard deviation of $48$ \textmu m (Fig.~\ref{fig:axial_resolution}(d)) is consistent with the actual displacement of $180$ \textmu m.

\begin{figure}[htbp]
	\centering
	\includegraphics[width=\linewidth]{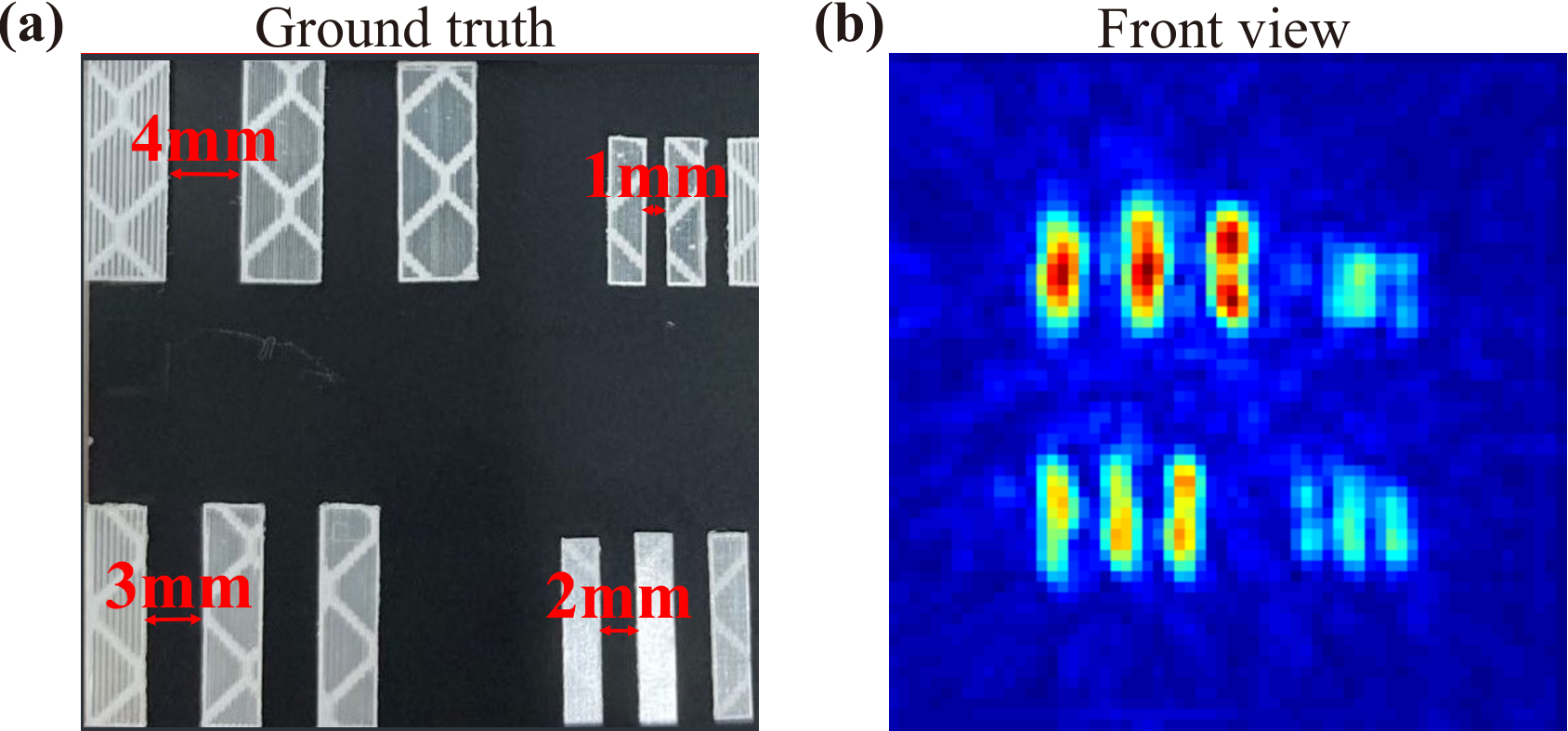}
	\caption{Lateral spatial resolution. (a) The hidden resolution chart used for the lateral resolution test. The intervals between strips in the four groups are $ 4 $ mm, $ 3 $ mm, $ 2 $ mm, and $ 1 $ mm respectively. (b) The reconstructed front view of the resolution chart with $ 64 \times 64 $ sampling points on the visible wall. }
	\label{fig:lateral_resolution}
\end{figure}

We use a resolution chart consisting of four groups of strips to measure the lateral resolution (Fig.~\ref{fig:lateral_resolution}(a)). The delay line is scanned with a step size of $75$ \textmu m (corresponding to $0.5$ ps) and dwelling for $50$ ms on each time bin. After the reconstruction with the LCT algorithm, we use the Laplacian of Gaussian (LOG) method for image enhancement. However, by setting the $1$ mm interval on the wall during the reconstruction, an image with obvious aberrations is obtained. Although the $1$ mm scanning step of the sampling points on the visible wall is set, there is a slight distortion mainly caused by the two-way mirror deflection scanning system and the unevenness of the delay line. With an optimized interval setting of $1.05$ mm on the visible wall for the reconstruction (see Figure S3 in the Supplemental Material \cite{supp}), the best result with a distinguishable $2$ mm gap (Fig.~\ref{fig:lateral_resolution}(b)) is obtained.

\begin{figure}[htbp]
	\centering
	\includegraphics[width=\linewidth]{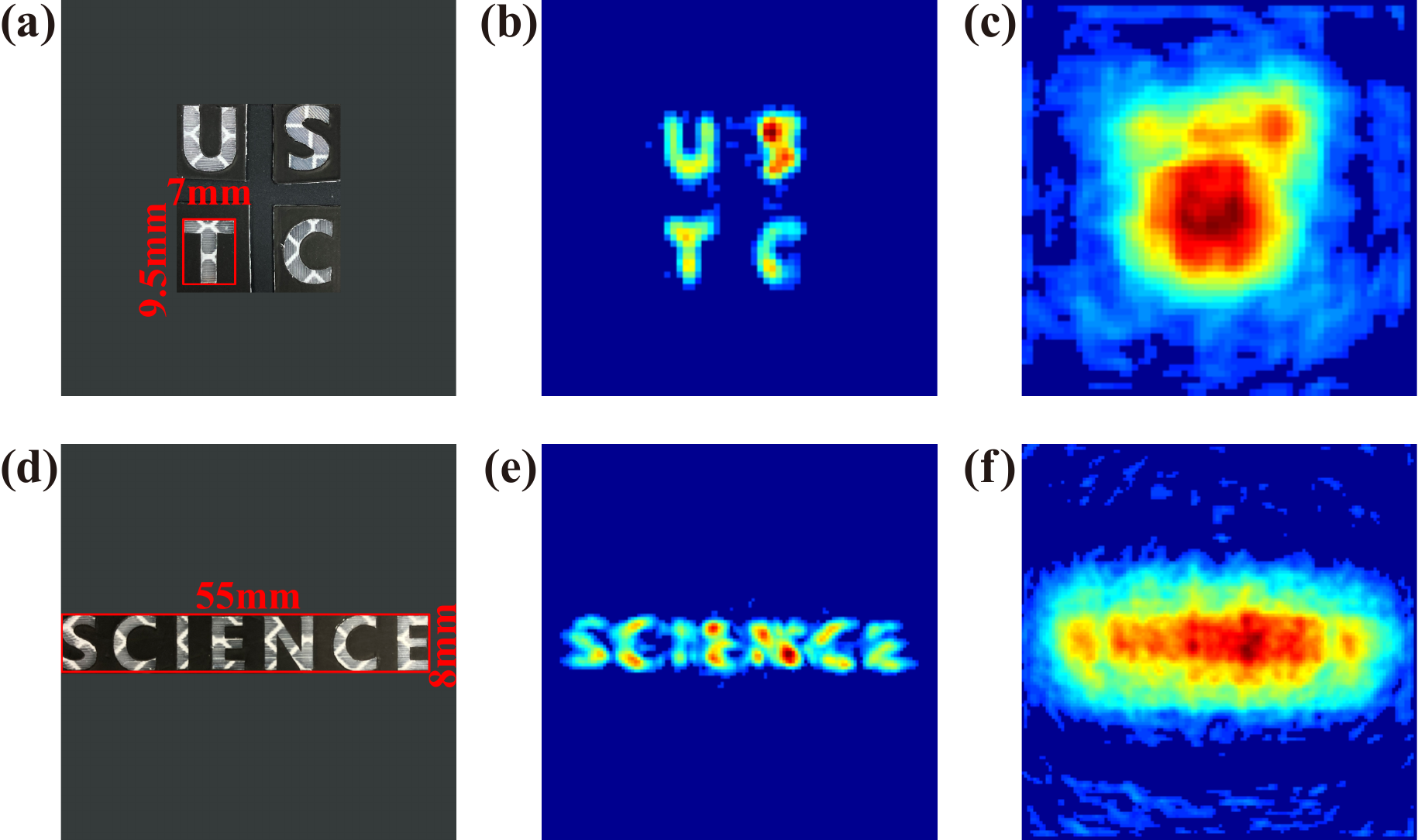}
	\caption{Reconstruction of letters. (a) Hidden 'USTC' letters. (b) Front view of the reconstructed image of 'USTC' with $ 64 \times 64 $ sampling points on the visible wall and $ 0.5 $ ps scanning time bin on the delay line. (c) Front view of the reconstructed image of 'USTC' with $ 5 $ ps temporal resolution. (d) Hidden word of 'SCIENCE'. (e) Front view of the reconstructed image of 'SCIENCE' with $ 128 \times 128 $ sampling points on the visible wall and $ 0.5 $ ps scanning time bin on the delay line. (f) Front view of the reconstructed image of 'SCIENCE' with $ 5 $ ps temporal resolution. }
	\label{fig:letters}
\end{figure}

With a $2$ mm lateral resolution, the ability to read hidden words was shown by imaging the letters 'USTC' and the word 'SCIENCE'. The heights of the 'USTC' letters are $ 9.5 $ mm (Fig.~\ref{fig:letters}(a)), and the height of the word 'SCIENCE' is $ 8 $ mm (Fig.~\ref{fig:letters}(d)). The letters 'USTC' and the word 'SCIENCE' have the same size of $ 36 $ pt and $ 30 $ pt, respectively, in bold Microsoft YaHei font printed on an A4 paper. We use $64 \times 64$ and $128 \times 128$ grid sampling on the visible wall for the 'USTC' and 'SCIENCE' hidden targets, respectively. The four letters in the first target can be clearly distinguished after the reconstruction with the LCT algorithm and sharpened by LOG filtering (Fig.~\ref{fig:letters}(b)). In Fig.~\ref{fig:letters}(e), we can tell most parts of the word 'SCIENCE'. Although we use a retro-reflective target, there are some Lambertian plastic areas (white lines), as can be seen in the actual images (Fig.~\ref{fig:letters}(a) and ~\ref{fig:letters}(d)). These Lambertian areas split some letters. Without such splitting, the word 'SCIENCE' could be clearly read as well. It should be noticed that the acquired images contain 3D information. The orientation of the targets (pitch and yaw angles) can be acquired from the side and top view of the targets (see Section 5 in the Supplemental Material \cite{supp}) and show the high axial resolution at the same time.

The picosecond temporal resolution in our system is critical to achieving NLOS imaging of such small hidden targets. This is proved by comparing the results obtained with a high and poor temporal resolution. By merging 10 time bins of the raw data into one single time bin, one may obtain the data with a $5$ ps temporal resolution. However, the reconstructed images with $5$ ps temporal resolution (Fig.~\ref{fig:letters}(c) and ~\ref{fig:letters}(f)) contain no information about the letters. This clearly shows the significance of the jitter of the imaging system in transient NLOS imaging, and the spatial resolution ability achieved here cannot be accomplished by an NLOS imaging system based on standard SPDs \cite{Hadfield2009NP,Zhang2015InGaAs,korzh2020psSNSPD}.

\emph{Conclusion.---}
Using a synchronized two-channel picosecond pulse laser, we construct a picosecond temporal resolution SPD for NLOS imaging. We verify a $180$ \textmu m axial resolution and a $\sim 2$ mm lateral resolution which is one order of magnitude improvement compared with the earlier reported results \cite{O'Toole2018LCT}. Furtherly, we realize high-resolution NLOS imaging. Simultaneous optimization of the detection efficiency, dark counts, and temporal resolution provides suitable conditions for NLOS imaging, and this is the first successful demonstration of using an up-conversion setup in the context of NLOS imaging. Furthermore, the recent development of the NLOS imaging reconstruction algorithms \cite{O'Toole2018LCT,O'Toole2019f-k,Velten2019PhasorCamera} gives full play to the high temporal resolution. Our work paves the way for a detailed reconstruction of hidden scenes in inaccessible areas, not only making it possible to read hidden words smaller than those demonstrated in previous work using a conventional camera \cite{Heide2020CNN}, but also providing depth information with a sub-millimeter resolution. The depth information is crucial. Considering complex hidden targets that are tilted with respect to the visible wall, the limited information offered by conventional cameras may not help us in distinguishing the target. Additionally, there is no FOV limitation in the hidden region compared to high-resolution NLOS imaging with a conventional camera \cite{Heide2020CNN}. In the future, the theoretical resolution limit may be achieved by improving the precision of positioning on the wall or adapting the reconstruction algorithm to a non-grid sampling point distribution on the visible wall. Moreover, the lateral resolution is also determined by the ratio between the position of the hidden objects and the scanning area on the wall: $z/w$. Therefore, by doubling the scanning area on the wall in our system, $\Delta x$ can be easily decreased to 0.6 mm, which is sufficient resolution for clear reading of out-of-sight handwriting. As the temporal resolution of SPD technology is constantly enhanced \cite{korzh2020psSNSPD}, NLOS imaging with 3D information will offer us an unprecedented detailing of hidden objects in different applications.

\begin{acknowledgments}
We thank Zheng-Ping Li, Jian-Jiang Liu and Cheng Wu for the helpful discussions. This work was supported by the National Key R\&D Program of China (2018YFB0504300), National Natural Science Foundation of China (61771443, 62031024), Key R\&D Plan of Shandong Province (2019JZZY010205), Natural Science Foundation of Shandong province (ZR2019LLZ003), Chinese Academy of Science, Shanghai Municipal Science and Technology Major Project (2019SHZDZX01), SAICT Experts Program, Taishan Scholar Program of Shandong Province, Quancheng Industrial Experts Program, 5150 Program for Talents Introduction and Key R\&D Program of Guangdong Province (2018B030325001).
\end{acknowledgments}


\end{document}